\documentclass[pra,aps,preprintnumbers,twocolumn,notitlepage,superscriptaddress,amsmath,amssymb]{revtex4-1}

\usepackage                             {cmap} 
\usepackage     [utf8]                  {inputenc}
\usepackage     [T1]                    {fontenc}
\usepackage                             {color}
\usepackage                             {amsmath}
\usepackage                             {graphicx}
\usepackage     [english]               {babel}
\usepackage                             {hyperref}
\usepackage	[usenames,dvipsnames]	{pstricks}
\usepackage				{dcolumn}
\usepackage				{bm}
\usepackage				{booktabs}

\begin{document}

\title{Improvement of the Basis for the Solution of the Dirac Equation in Cassini Coordinates}

\author{Walter Hahn}
\email{w.hahn@thphys.uni-heidelberg.de}
\affiliation{Skolkovo Institute of Science and Technology, Skolkovo Innovation Centre, Nobel Street 3, Moscow 143026, Russia}
\affiliation{Institut f\"ur theoretische Physik, Universit\"at Heidelberg, Philosophenweg 19, 69120 Heidelberg, Germany}

\author{Anton N. Artemyev}
\email{anton.artemyev@physik.uni-kassel.de}
\affiliation{Institut f\"ur Physik, University of Kassel, Heinrich-Plett-Str. 40, 34132 Kassel, Germany}

\author{Andrey Surzhykov}
\email{andrey.surzhykov@ptb.de}
\affiliation{Physikalisch-Technische Bundesanstalt, D-38116 Braunschweig, Germany}
\affiliation{Technische Universit\"at Braunschweig, D-38106 Braunschweig, Germany}

\begin{abstract}
We propose an improvement of the basis for the solution of the stationary two-centre Dirac equation in Cassini coordinates using the finite-basis-set method presented in Ref.~\cite{pp}. For the calculations in Ref.~\cite{pp}, we constructed the basis for approximating the energy eigenfunctions by using smooth piecewise defined polynomials, called B-splines. In the present article, we report that an analysis of the employed representation of the Dirac matrices shows that the above approximation is not efficient using B-spines only. Therefore, we include basis functions which are defined  using functions with step-like behaviour instead of B-splines. Thereby, we achieve a significant increase of accuracy of results as compared to Ref.~\cite{pp}.
\end{abstract}

\maketitle

\section{Introduction}
Heavy highly-charged ions can serve as a tool for testing physical theories in the limit of extremely strong
electromagnetic fields. Of special interest, hereby, are slow
collisions of two highly-charged ions with the total charge $Z=Z_1+Z_2 > 173$, where $Z_1$ and $Z_2$ are the charges of the individual ions. If the
collision energy is about \mbox{$5$ MeV/u} and the impact parameter is close to
zero, the ions may form a short-living quasi-molecule during the fly by. Within the lifetime of such a quasi-molecule, the molecular ground
state dives into the negative continuum and, in this case, spontaneous pair creation is predicted theoretically. An experimental test of this prediction is planned at the future FAIR facility~\cite{fairdesign}. For an overview of theoretical and experimental
advances in this research field, we refer the reader to the following Refs.~\cite{pp,kozhedub,rumrich,gumb1,gumb2} and references therein.

In order to investigate the properties of these quasi-molecules, the Dirac equation for an electron in the potential of two moving nuclei must be solved. The following two approximations are usually made in such calculations. (1) The collision dynamics can be treated adiabatically due to the slow collision dynamics as compared with the average velocity of the bound electron. (2) The motion of the nuclei can be described classically by Rutherford trajectories because of the small ratio between the electron and the nuclear mass \cite{eichler}. With these two approximations, the
solution of the time-dependent Dirac equation can be traced back to that
of the stationary Dirac equation with two spatially fixed nuclei. We refer to the latter as the two-centre Dirac equation in the following and we use the notions of centre and nucleus interchangeably.

In our previous article~\cite{pp}, we proposed a novel method for the solution of the two-centre Dirac equation in Cassini coordinates. This method is based on the application of the finite-basis-set approach, where the basis was constructed using B-splines~\cite{deboor}. With this method, we obtained good approximations for the energy eigenstates and the energy eigenvalues of the problem. However, the convergence properties of the method varied with the distance between the two centres. For practical calculations, though, good convergence properties independent of the distance are necessary. Examples of such calculations are studies of ion collisions and the investigation of QED effects.

In the present article, we analyse the transformation of the Dirac equation from Cartesian to Cassini coordinates and the related transformation of the Dirac matrices~\cite{wietschlu1,wietschlu2}. We find that approximating the eigenfunctions in a finite basis which is constructed using B-splines only is not efficient. Therefore, we propose to extend the basis by a small number of basis functions which are constructed using functions with step-like behaviour instead B-splines. When including this proposal into our numerical routine, we achieve an improved accuracy of the method for all distances between the two centres.

The article is organised as follows. In Sec.~2, we provide an outline of the method used in Ref.~\cite{pp}. In Sec.~3, we describe our proposal for the improvement of this method. In Sec.~4, we demonstrate the improvement of the calculated results for the one-centre problem. Finally, a brief summary is given in Sec.~5. We use natural units ($m_e=c=\hbar=1$) throughout the paper.

\section{Description of the method used in Ref.~\cite{pp}}\label{secprelim}
Let us first consider the Dirac equation for an electron in the potential created by two spatially fixed nuclei. In the Cartesian coordinates, this Dirac equation reads ${\cal H}\Psi(x,y,z)=E\Psi(x,y,z)$ with
\begin{equation} \label{dir_ham}
 {\cal H}=-i\left(\alpha_x\frac{\partial}{\partial x}+\alpha_y\frac{\partial}{\partial y}+\alpha_z\frac{\partial}{\partial z}\right)+V(x,y,z)+\beta,
\end{equation}
where $\alpha_x$, $\alpha_y$, $\alpha_z$ and $\beta$ are the Dirac matrices and $V(x,y,z)$ is the two-centre potential. This potential can be written as $V(x,y,z)=V_1(x,y,z)+V_2(x,y,z)$ with $V_i(x,y,z)$ being the potential created by the $i$-th nucleus.

\subsection{Cassini coordinates and the transformation of the Dirac equation}\label{cassss}
For the solution of the above two-centre problem, we choose the Cassini coordinates which are defined as follows:
\begin{equation}\label{casscoord}
 w\equiv\frac{\sqrt{r_1r_2}}{a},\ \ \ \ \ \ \ \ \ \ \delta\equiv\frac{\theta_1+\theta_2}{2},\ \ \ \ \ \ \ \ \ \ \phi\equiv\phi,
\end{equation}

\begin{figure}[b]
 \includegraphics[width=7.5cm]{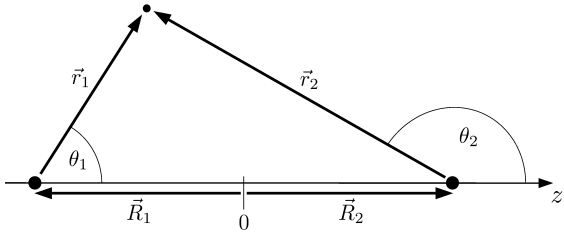}
 \caption{Illustration of the notations used for the definition of the Cassini coordinates in Eq.~(\ref{casscoord}). See text for further explanations.}
 \label{piccassini_1}
\end{figure}
\begin{figure}
  \includegraphics[width=5cm]{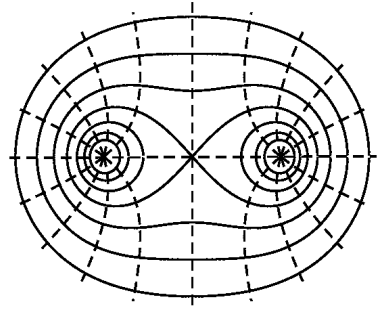}
  \caption{Lines of constant values of $w$ (solid lines) and $\delta$ (dashed lines).} 
\label{piccassini_2}
\end{figure}


where $r_i\equiv|\vec{r}_i|$ is the distance between the electron and the $i$-th
nucleus, \mbox{$a\equiv|\vec{R}_1-\vec{R}_2|/2$} is a half of the distance
between the nuclei (cf. Fig.~\ref{piccassini_1}), $\theta_i$ is the angle between the internuclear axis
($z$-axis in Fig.~\ref{piccassini_1}) and the vector $\vec{r}_i$, and $\phi$ is
the azimuthal angle. Among coordinate systems suitable for two-centre problems with cylindrical
symmetry, Cassini coordinates are particularly useful because surfaces
of constant values of $w$, cf. Fig.~\ref{piccassini_2}, almost coincide with equipotential
surfaces of two point-like nuclei of the same charge~\cite{wietschlu2}. Since this near-coincidence is largely independent of the internuclear distance, Cassini coordinates are suitable for both small and large distances between the nuclei.

In order to facilitate the numerical treatment of the Dirac Hamiltonian~(\ref{dir_ham}), we choose the representation of Dirac matrices introduced in Refs.~\cite{wietschlu1,wietschlu2}. In this representation, the Dirac equation in Cassini coordinates reads ${\cal H}_\textnormal{\scriptsize Cassini}\Psi(w,\delta,\phi)=E\Psi(w,\delta,\phi)$ with
\begin{eqnarray}\label{transhamil2}
 {\cal H}_\textnormal{\scriptsize Cassini}=-i\frac{D^{1/4}}{aw}\left(\alpha_3\frac{\partial}{\partial
   w}+\alpha_1\frac{1}{w}\frac{\partial}{\partial\delta}\right)-i\alpha_2\frac{1}{\rho}\frac{\partial}{\partial\phi}\\
   +V(w,\delta;a)+\beta,\nonumber
\end{eqnarray}
where $\alpha_1$, $\alpha_2$, $\alpha_3$ and $\beta$ are Dirac matrices, $V(w,\delta;a)$ is the two-centre potential written in Cassini coordinates, $\rho$ is the distance to the internuclear axis 
\begin{equation} \label{eqn_rho}
 \rho=\frac{a}{\sqrt{2}}\sqrt{\sqrt{D}-1-w^2\cos2\delta}
\end{equation}
and
\begin{equation}\label{rhod}
  D=w^4+2w^2\cos2\delta+1.
\end{equation}
We choose the following representation of the Dirac matrices~\cite{wietschlu2}
\begin{eqnarray}\label{repres_1}
 &&\alpha_1\!=\!\left(\!\!\!\begin{array}{cc} 0 & \sigma_2 \\ \sigma_2 &
 0 \end{array}\!\!\!\right)\!,\ \ \alpha_2\!=\!\left(\!\!\!\begin{array}{cc} 0 &
 \sigma_1 \\ \sigma_1 & 0 \end{array}\!\!\!\right)\!,\\
 &&\alpha_3\!=\!\left(\!\!\!\begin{array}{cc} 0 & -iI \\ iI &
 0 \end{array}\!\!\!\right)\!,\ \ \beta\!=\!\left(\!\!\!\begin{array}{cc} I & 0
 \\ 0 & I \end{array}\!\!\!\right)\!, \label{repres_2}
\end{eqnarray}
where $\sigma_1$ and $\sigma_2$ are the Pauli matrices, $0$ and $I$ are the 
$2\times 2$ zero and unit matrix, respectively.

The wave function $\Psi(x,y,z)$ in the Cartesian coordinates is related to the above wave function $\Psi(w,\delta,\phi)$ in Cassini coordinates by \cite{wietschlu2} 
\begin{equation}\label{wavetrans}
 \Psi(w,\delta,\phi)=\sqrt{\rho}\left(\frac{w^2a}{\sqrt[4]{D}}\right)^{1/2}\textbf{S}^{-1}_2(\alpha)\textbf{S}^{-1}_3(\phi)\,\Psi(x,y,z), 
\end{equation}
where $\textbf{S}_3(\phi)=\exp\left(-\frac{i}{2}\phi\Sigma_3\right)$ and
$\textbf{S}_2(\alpha)=\exp\left(-\frac{i}{2}\alpha\Sigma_2\right)$ with $\Sigma_2$
and $\Sigma_3$ being constant $4\times 4$ matrices explicitly given in Ref.~\cite{wietschlu1}. In Eq.~(\ref{wavetrans}), the function $\alpha(w,\delta)$ is defined by
\begin{equation}\label{cosfactor}
 \cos(\alpha)=\frac{\sqrt{\sqrt{D}+1+w^2\cos(2\delta)}\left(\sqrt{D}-1\right)}{\sqrt{2}w^2\sqrt[4]{D}}\ \textrm{sgn}[\cos(\delta)]
\end{equation}
and
\begin{equation}\label{sinfactor}
 \sin(\alpha)=\frac{\sqrt{\sqrt{D}-1-w^2\cos(2\delta)}\left(\sqrt{D}+1\right)}{\sqrt{2}w^2\sqrt[4]{D}},
\end{equation}
where $\textrm{sgn}[\cos(\delta)]$ denotes the sign of $\cos(\delta)$. We refer to the term in front of $\Psi(x,y,z)$ in Eq.~(\ref{wavetrans}) as the transformation matrix in the following.

Due to the axial symmetry of the problem with respect to the internuclear axis, the projection of the total angular momentum $\mu$ on the internuclear axis is conserved. Therefore, the eigenfunctions of the Hamiltonian~(\ref{transhamil2}) can be written of the form $\Psi(w,\delta,\phi)=\psi_\mu(w,\delta)\exp(i\mu\phi)$, where $\mu=\pm1/2, \pm3/2, ...$. Therefore, we can rewrite the Hamiltonian~(\ref{transhamil2}) as
\begin{eqnarray} \label{cas_ham}
 {\cal H}^{(\mu)}_\textnormal{\scriptsize Cassini}=-i\frac{D^{1/4}}{aw}\left(\alpha_3\frac{\partial}{\partial
   w}+\alpha_1\frac{1}{w}\frac{\partial}{\partial\delta}\right)+\alpha_2\frac{\mu}{\rho}\\
   +V(w,\delta;a)+\beta.\nonumber
\end{eqnarray}
Given the representation of the Dirac matrices in Eq.~(\ref{repres_1}) and Eq.~(\ref{repres_2}), ${\cal H}^{(\mu)}_\textnormal{\scriptsize Cassini}$ is real. Since $\alpha_1$, $\alpha_2$, $\alpha_3$, and $\beta$ are hermitian, ${\cal H}^{(\mu)}_\textnormal{\scriptsize Cassini}$ is also hermitian.

\subsection{Finite-basis-set approach and B-splines}\label{secfinit}
To find approximate expressions for the eigenfunctions and the eigenvalues of ${\cal H}^{(\mu)}_\textnormal{\scriptsize Cassini}$ in Eq.~(\ref{cas_ham}), we rewrite the above wave function $\psi_\mu(w,\delta)$ as a linear combination of $N_\textnormal{\scriptsize f}$ four-component functions $u_k(w,\delta)$
\begin{equation}\label{finiteexpans}
 \psi_\mu(w,\delta)\approx\sum_{k=1}^{N_\textnormal{\scriptsize f}}C_{\mu k} u_k(w,\delta),
\end{equation}
where $C_{\mu k}$ are the expansion coefficients. We choose the functions $u_k(w,\delta)$ to be square integrable and to satisfy the boundary conditions of the problem. In general, the larger the number $N_\textnormal{\scriptsize f}$ of the functions $u_k(w,\delta)$, the better is the approximation~(\ref{finiteexpans}). We refer to the set of the functions $u_k(w,\delta)$ as basis in the following.

The variational principle leads to the following generalised eigenvalue problem
\begin{equation}\label{seteq}
 \sum_{k=1}^{N_\textnormal{\scriptsize f}} A_{ik}C_{\mu k}=\varepsilon\sum_{k=1}^{N_\textnormal{\scriptsize f}}  K_{ik}C_{\mu k},
\end{equation}
where $A_{ik}=[\langle u_i|{\cal H}^{(\mu)}_\textnormal{\scriptsize Cassini}|u_k \rangle +\langle
u_k|{\cal H}^{(\mu)}_\textnormal{\scriptsize Cassini}|u_i\rangle ]/2$, $K_{ik}=\langle u_i|u_k
\rangle$ and $\varepsilon$ is the energy eigenvalue. Due to the non-orthogonality of the functions $u_k(w,\delta)$, the matrix $K_{ik}$ in Eq.~(\ref{seteq}) differs from the unit matrix. The solution of this eigenvalue problem yields the energy eigenvalues and eigenfunctions of the problem.

Following Ref.~\cite{pp}, we define the functions $u_k(w,\delta)$ as
\begin{equation}\label{strucbas} 
 u_{k}(w,\delta)= b_{k}\ {\cal B}_{k}(w,\delta),
\end{equation}
where $b_{k}$ is one of the following four-component vectors
\begin{equation} \label{bb}
b_k=\left(\begin{array}{c}
 1 \\
 0 \\
 0 \\
 0
 \end{array}\right),\ \ 
b_k=\left(\begin{array}{c}
 0 \\
 1 \\
 0 \\
 0
 \end{array}\right),\ \ 
 b_k=\left(\begin{array}{c}
 0 \\
 0 \\
 1 \\
 0
 \end{array}\right),\ \ 
 b_k=\left(\begin{array}{c}
 0 \\
 0 \\
 0 \\
 1
 \end{array}\right),
\end{equation}
and ${\cal B}_{k}(w,\delta)$ is one of $N$ scalar functions which are of the form
\begin{equation}\label{scalarstruc}
{\cal B}_{k}(w,\delta)=f(w,\delta)B_{k}^{n_w}(w)B_{k}^{n_\delta}(\delta),
\end{equation}
where $f(w,\delta)\equiv\sqrt{\rho}/\sqrt[4]{D}$, and $B_{k}^{n_w}(w)$ and $B_{k}^{n_\delta}(\delta)$ are B-splines of the order $n_w$ and $n_\delta$, respectively~\cite{deboor}. The function $f(w,\delta)$ in Eq.~(\ref{scalarstruc}) is introduced to account for non-polynomial behaviour of the wave function. For the definition of the functions $u_k(w,\delta)$ in~(\ref{strucbas}), each of the four vectors $b_k$ in Eq.~(\ref{bb}) is combined with each scalar function ${\cal B}_{k}(w,\delta)$. Therefore, $N_\textnormal{\scriptsize f}=4N$.

\section{Proposal for the improvement of the set of scalar functions ${\cal B}_k(w,\delta)$} \label{secref}
In Ref.~\cite{pp}, we obtained the eigenvalues and the eigenfunctions of ${\cal H}^{(\mu)}_\textnormal{\scriptsize Cassini}$ in Eq.~(\ref{cas_ham}) by using the finite-basis-set method together with the basis functions $u_{k}(w,\delta)$ as described above. In order to determine the accuracy of this method, we applied the method to the one-centre Dirac problem by setting the charge of one nucleus to zero and then compared the calculated energy
levels with the well known levels of hydrogen-like ions (see, for example, Ref.~\cite{art_comp}). We found that the method reproduces the level structure very accurately. However, the accuracy of the obtained results varied with the inter-centre distance $2a$.

In this section, we propose a modification of the above method in order to improve the accuracy of results for all inter-centre distances. We show that due to the properties the transformation matrix in Eq.~(\ref{wavetrans}) the approximation~(\ref{finiteexpans}) is not efficient using the scalar functions defined in Eq.~(\ref{scalarstruc}).

\subsection{Properties of the transformation matrix in Eq.~(\ref{wavetrans})}
\begin{figure}[t]
 \centering
 \includegraphics[width=8cm]{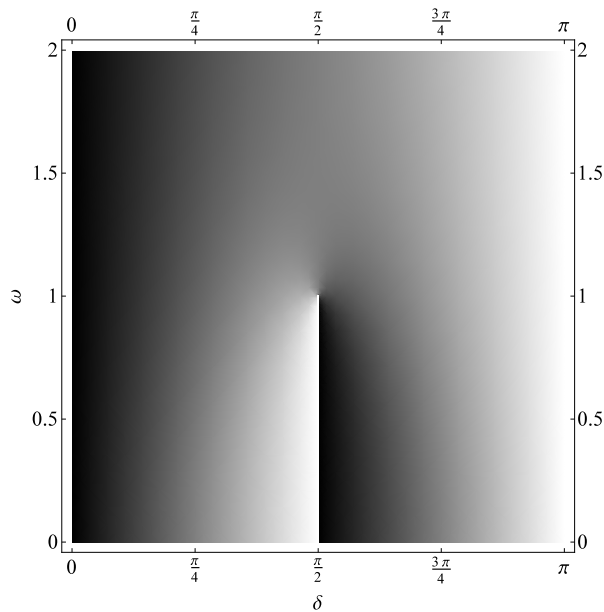}
 \caption{\small A grey-scale plot of the function $\alpha(w,\delta)$ defined in Eq.~(\ref{cosfactor}) and Eq.~(\ref{sinfactor}) for the interval $0\leq\delta\leq\pi$ and $0\leq w\leq2$, where the value of $\alpha(w,\delta)$ is encoded by different shades of grey. Black corresponds to $\alpha(w,\delta)=0$ and white corresponds to $\alpha(w,\delta)=\pi$. The figure shows that, for $w>1$, $\alpha(w,\delta)$ is continuous, whereas, for $w<1$, $\alpha(w,\delta)$ jumps by $\pi$ at $\delta=\pi/2$.} 
 \label{pic_discont}
\end{figure}

Let us focus on the transformation matrix introduced in Eq.~(\ref{wavetrans}) and, in particular, investigate the function $\alpha(w,\delta)$ defined in Eq.~(\ref{cosfactor}) and Eq.~(\ref{sinfactor}). It turns out that the function $\alpha(w,\delta)$ is not continuous along the internuclear line which is defined by $\delta=\pi/2$ and $0<w<1$, cf. Fig.~\ref{piccassini_1}. We obtain for all $w<1$
\begin{equation}
 \lim_{\delta\nearrow\pi/2}\alpha(w,\delta)=\pi\quad\textnormal{and}\ \lim_{\delta\searrow\pi/2}\alpha(w,\delta)=0,
\end{equation}
where $\delta\!\nearrow\!\pi/2$ denotes the limit $\delta\to\pi/2$ from below, while $\delta\!\searrow\!\pi/2$ denotes the same limit from above. This discontinuity is due to the factor $\textrm{sgn}[\cos(\delta)]$ in Eq.~(\ref{cosfactor}). In Figure~\ref{pic_discont}, we show a grey-scale plot of $\alpha(w,\delta)$ in the interval $0\leq\delta\leq\pi$ and $0\leq w\leq2$, where the value of $\alpha(w,\delta)$ is encoded by the different shades of grey. The colour black corresponds to $\alpha(w,\delta)=0$ and the colour white corresponds to $\alpha(w,\delta)=\pi$. While $\alpha(w,\delta)$ is continuous for $w>1$, it makes a jump at $\delta=\pi/2$ for all $w<1$.

From the above discontinuity of $\alpha(w,\delta)$ follows that the term $\textbf{S}^{-1}_2(\alpha)$ of the transformation matrix in Eq.~(\ref{wavetrans}) is also not continuous along the internuclear line. However, the wave function $\Psi(w,\delta,\phi)$ itself is continuous because of the factor $\sqrt{\rho}$ in Eq.~(\ref{wavetrans}), where $\sqrt{\rho}=0$ along the internuclear line, cf. Eq.~(\ref{eqn_rho}). Let us now consider the approximation of the wave function $\Psi(w,\delta,\phi)$ in Eq.~(\ref{finiteexpans}). The factor $\sqrt{\rho}$, which makes $\Psi(w,\delta,\phi)$ continuous as discussed above, is accounted for by the function $f(w,\delta)$ in the definition of ${\cal B}_{k}(w,\delta)$ in Eq.~(\ref{scalarstruc}). Therefore, it is instructive to consider the function $\Psi(w,\delta,\phi)/\sqrt{\rho}$. This function is not continuous along the internuclear line. Comparing now this function with the expression $B_{k}^{n_w}(w)B_{k}^{n_\delta}(\delta)/\sqrt[4]{D}$, cf. Eq.~(\ref{scalarstruc}), which we use for the approximation of $\Psi(w,\delta,\phi)/\sqrt{\rho}$, it turns out that we approximate a discontinuous function by continuous \mbox{B-splines}. This is not an efficient approximation.

\subsection{Definition of additional functions}
In order to account for the above discontinuity of $\Psi(w,\delta,\phi)/\sqrt{\rho}$ in our numerical treatment, we add to the set of scalar functions defined in Eq.~(\ref{scalarstruc}) the following functions: For each B-spline $B_{k}^{n_w}(w)$, which vanishes outside the interval $w\in[0,1+\epsilon]$ with $\epsilon\ll1$, we define the scalar function
\begin{equation}
 {\cal B}'_{k}(w,\delta)\equiv f(w,\delta)B_{k}^{n_w}(w)F(\delta),
\end{equation}
where $F(\delta)$  is a function, which is discontinuous at \mbox{$\delta=\pi/2$}, and $f(w,\delta)$ is defined as in the previous section. We define the new basis functions as \mbox{$u'_{k}(w,\delta)=b_k{\cal B}'_{k}(w,\delta)$} by combining ${\cal B}'_{k}(w,\delta)$ with each of the vectors $b_k$ in Eq.~(\ref{bb}) and we refer to these functions as additional functions. We achieved best results with the Heaviside-step-function
\begin{equation}
 F(\delta)=\Theta\left(\delta-\frac{\pi}{2}\right).
\end{equation}
We use this expression for $F(\delta)$ throughout this article.

\section{Calculations for the one-centre problem as a benchmark}\label{resstan}
In this section, we demonstrate the improvement of the method due to the additional functions. We set the charge of one of the nucleus to zero and calculate the low-energy states of hydrogen-like Uranium with and without the additional functions. We also compare the results with known high-accuracy results (see, for example, Ref.~\cite{art_comp}). The calculated energy eigenvalues parametrised by the principal $n$ and the angular $\kappa$ quantum numbers are shown in Table~\ref{tabl3} for different values of the inter-centre distance $d\equiv2a$.

\begin{table*}[t]
\begin{center}\small
\begin{tabular}[b]{llrcccc}
 \toprule
  & & & \multicolumn{2}{c}{no additional functions} & \multicolumn{2}{c}{with additional functions} \\
 \cmidrule(r){4-5} \cmidrule(r){6-7}
d [a.u.] & n & $\kappa$ & E [mc$^2$] & rel. accuracy & E [mc$^2$] & rel.
 accuracy \\
 \toprule
 $10^{-2}$   & 1 & -1 & 0.74140000 & $2\times 10^{-4}$ & 0.74153043 & $1\times 10^{-5}$\\
 \cmidrule(r){2-7}
  & 2 & -1 & 0.93311589 & $2\times 10^{-7}$ & 0.93311577& $1\times 10^{-7}$\\
 \cmidrule(r){2-7}
& 2 & 1 & 0.93305372 & $3\times 10^{-6}$ & 0.93304967 & $1\times 10^{-6}$\\
 \cmidrule(r){2-7}
  & 2 & -2 & 0.94197617 & $6\times 10^{-7}$ & 0.94197678 & $5\times 10^{-8}$\\
 \toprule
 $10^{-1}$   & 1 & -1 & 0.74133016 & $3\times 10^{-4}$ & 0.74152243 & $5\times 10^{-7}$\\
 \cmidrule(r){2-7}
  & 2 & -1 & 0.93320997 & 2$\times 10^{-4}$ & 0.93311330& $3\times 10^{-6}$\\
 \cmidrule(r){2-7}
& 2 & 1 & 0.93302997 & $2\times 10^{-5}$ & 0.93304775 & $3\times 10^{-6}$\\
 \cmidrule(r){2-7}
  & 2 & -2 & 0.94187430 & $1\times 10^{-4}$ & 0.94197567 & $1\times 10^{-6}$\\
 \toprule
 $10^{+0}$   & 1 & -1 & 0.74135637 & $2\times 10^{-4}$ & 0.74152852 & $8\times 10^{-6}$\\
 \cmidrule(r){2-7}
  & 2 & -1 & 0.93309642 & 2$\times 10^{-5}$ & 0.93311679& $1\times 10^{-6}$\\
 \cmidrule(r){2-7}
& 2 & 1 & 0.93302849 & $2\times 10^{-5}$ & 0.93305073 & $2\times 10^{-7}$\\
 \cmidrule(r){2-7}
  & 2 & -2 & 0.94193602 & $4\times 10^{-5}$ & 0.94197421 & $3\times 10^{-6}$\\
 \toprule
 $10^{+1}$   & 1 & -1 & 0.74135455 & $2\times 10^{-4}$ & 0.74152409 & $2\times 10^{-6}$\\
 \cmidrule(r){2-7}
  & 2 & -1 & 0.93309207 & $3\times 10^{-5}$ & 0.93311254& $3\times 10^{-6}$\\
 \cmidrule(r){2-7}
& 2 & 1 & 0.93302645 & $3\times 10^{-5}$ & 0.93304811 & $3\times 10^{-6}$\\
 \cmidrule(r){2-7}
  & 2 & -2 & 0.94193303 & $5\times 10^{-5}$ & 0.94197698 & $3\times 10^{-7}$\\
 \toprule
 $10^{+2}$   & 1 & -1 & 0.74135485 & $2\times 10^{-4}$ & 0.74152409 & $2\times 10^{-6}$\\
 \cmidrule(r){2-7}
  & 2 & -1 & 0.93309216 & $3\times 10^{-5}$ &0.93311255 & $3\times 10^{-6}$\\
 \cmidrule(r){2-7}
& 2 & 1 & 0.93302651 & $3\times 10^{-5}$ & 0.93304813 & $3\times 10^{-6}$\\
 \cmidrule(r){2-7}
  & 2 & -2 & 0.94193252 & $5\times 10^{-5}$ & 0.94197702 & $3\times 10^{-7}$\\
 \bottomrule
\end{tabular}

\caption{\small Calculated energies of the low-lying bound states of hydrogen-like Uranium for different inter-centre distances $d=2a$ in atomic units. The second and third columns indicate principal $n$ and angular $\kappa$ quantum numbers of the
state if it is considered in spherical coordinates. The fourth and the sixth
columns provide the values of the energy calculated without
and with additional functions, respectively. The fifth and seventh
columns show the relative accuracy of the obtained values with respect to calculations in a spherical coordinate system (see, for example, Ref.~\cite{art_comp}). Numerical values of fundamental constants used in the calculations were $\alpha^{-1}=137.0359895$ and $hcR_{\infty}=13.6056981$eV. The root-mean-square radius of the nucleus was $R=5.8507$fm, which was taken from Ref.~\cite{rms}.}
\label{tabl3}
\end{center}
\end{table*}

The parameters for the calculations were as follows. The projection of the total angular momentum was \mbox{$\mu=1/2$}. For the nuclear charge density, we used a two-parameter Fermi distribution which is explicitly given in Ref.~\cite{pp}. For the integration and the solution of the generalised eigenvalue problem~(\ref{seteq}), we used the same algorithms as in Ref.~\cite{pp}. The distributions of knots for the definition of B-splines were as follows: The distribution of $w$-knots was equidistant for $w<1$ and exponential for $w>1$, and the distribution of $\delta$-knots was equidistant. Additional knots around the point $w=1$, $\delta=\pi/2$ were inserted.

The accuracy of the calculated energy eigenvalues increases for all distances after including additional functions. This
increase of accuracy holds for almost all low-energy eigenstates. In total, the accuracy is
good compared to the relatively small number of basis functions used in our
calculations, such that the computation time is of the order of a few minutes on a
standard personal computer. The total number of basis functions was about 540
for $a<10$ and 450 for larger $a$. The number of additional functions varied
from 16 to 30 which corresponds to an increase of the number of basis functions by less than 7\% but, in fact, leads to an increase of accuracy by at least one order of magnitude for the states shown in Table~\ref{tabl3}. These additional functions are the only difference for the calculated values. We also obtain similar results for other hydrogen-like ions.

Non-physical spurious states, which typically appear in applications of the variational principle to relativistic problems~\cite{shabaevdkb,canada}, were identified by means of their characte\-ristic oscillatory behaviour and were excluded in Table~\ref{tabl3}.

The question whether our proposal is useful for the solution of the two-centre problem~\cite{chargetrans,kulkol,mironova} can be resolved by practical calculations for this problem which is left for further studies.

\section{Conclusions and outlook}
In this article, we described a proposal for the improvement of the solution of the stationary two-centre Dirac equation in Cassini coordinates using the finite-basis-set method. An analysis of the transformation of the Dirac matrices showed that the approximation of the wave function $\Psi(w,\delta,\phi)$ in Cassini coordinates as described in Section~2 is not efficient using B-splines only. Therefore, we included additional functions which are defined using functions with step-like behaviour instead of B-splines and, thereby, achieved a significant improvement of the convergence properties of the method.

When using our proposal in practical calculations, a given precision for the energy eigenstates and eigenvalues can be achieved with a smaller basis as compared to calculations without using our proposal. This is, for example, important for calculations of the structure properties (including QED) and dynamics of quasi-molecules. A further improvement could be possibly made by using the dual kinetic-balance basis~\cite{shabaevdkb} in order to avoid spurious states.

\textit{Acknowledgements:} The work is supported by the Helmholtz Gemeinschaft under the project VH-NG-421. W.H. is grateful for the support from the \textit{Studienstiftung des deutschen Volkes}.


\end{document}